\documentclass[aps,pra,showpacs,amsmath,amssymb,superscriptaddress,reprint,10pt,longbibliography]{revtex4-1}
\usepackage{graphicx,mathrsfs,enumerate}
\usepackage[utf8]{inputenc}
\usepackage{mathtools}
\usepackage{amsfonts,amstext}
\usepackage{amsmath,amsthm,amssymb,mathbbol,tensor}
\usepackage{graphicx,xcolor,mathrsfs,enumerate}
\usepackage[hidelinks]{hyperref}\hypersetup{pdfpagemode=UseNone,pdfstartview=}
\usepackage[left=2cm,right=2cm]{geometry}
\usepackage{url}
\usepackage{times}
\usepackage{bbm}
\usepackage[english]{babel}

\newcommand{\bra}[1]{\langle #1|}
\newcommand{\ket}[1]{|#1\rangle}

\newcommand{\ketbra}[2]{| #1 \rangle \langle #2 |}

\DeclareMathOperator{\Tr}{Tr}

\newtheorem{thm}{Theorem}

\begin{document}
\title{Quantum steering beyond instrumental causal networks}
\author{R. V. Nery}
\affiliation{Instituto de F\'isica, Universidade Federal do Rio de Janeiro, P. O. Box 68528, Rio de Janeiro, RJ 21941-972, Brazil}
\author{M. M. Taddei}
\affiliation{Instituto de F\'isica, Universidade Federal do Rio de Janeiro, P. O. Box 68528, Rio de Janeiro, RJ 21941-972, Brazil}
\author{R. Chaves}
\affiliation{International Institute of Physics, Federal University of Rio Grande do Norte, 59078-970, P. O. Box 1613, Natal, Brazil}
\author{L. Aolita}
\affiliation{Instituto de F\'isica, Universidade Federal do Rio de Janeiro, P. O. Box 68528, Rio de Janeiro, RJ 21941-972, Brazil}
\affiliation{ICTP South American Institute for Fundamental Research
Instituto de F\'isica Te\'orica, UNESP-Universidade Estadual Paulista R. Dr. Bento T. Ferraz 271, Bl. II, S\~ao Paulo 01140-070, SP, Brazil}

\begin{abstract}
We theoretically predict, and experimentally verify with entangled photons, that outcome communication is not enough for hidden-state models to reproduce quantum steering.
Hidden-state models with outcome communication correspond, in turn, to the well-known \emph{instrumental processes}  of causal inference but in the 1-sided device-independent (1S DI) scenario of one black-box measurement device and one well-characterised quantum apparatus.
We introduce \emph{1S-DI instrumental inequalities} to test against these models, with the appealing feature of detecting entanglement even when communication of the black box's measurement outcome is allowed. 
We find that, remarkably, these inequalities can also be violated solely with steering, i.e. without outcome communication.
In fact, an efficiently-computable formal quantifier -- the \emph{robustness of non-instrumentality} -- naturally arises; and we prove that steering alone is enough to maximize it. Our findings imply that quantum theory admits a \emph{stronger form of steering} than known until now, with fundamental as well as practical potential implications.
\end{abstract}


\maketitle

\emph{Instrumental causal networks} are one of the main tools of causal inference \cite{Spirtes2001, Pearl2009}. Introduced almost a century ago \cite{Wright1928tariff} in the context of supply-and-demand models, they find nowadays a broad range of applications, from epidemiology and clinical trials \cite{BP97,Greenland2000} to econometrics \cite{Angrist1996} and ecology \cite{Creel2009}, e.g. In fact,  the instrumental causal structure  is special because it is the simplest one for which the strength of causal influences can be estimated solely from observational data -- i.e. without interventions -- even in the presence of hidden common causes \cite{Pearl1995}. 
Recently, considerable effort has been put into the quantisation of the classical theory of causality  \cite{Spirtes2001, Pearl2009}, giving rise to the so-called \emph{quantum causal networks} \cite{Chiribella2009,Oreshkov2012,Leifer2013,Henson2014,Chaves2015,Pienaar2015,Costa2016,Allen2016,Portmann2017}.
Apart from its implications in nonlocality \cite{Fritz2012,Chaves2015b,Chaves2016,Fritz2016, Rosset2016,Wolfe2016,CCA16}, the young field has brought about fascinating discoveries \cite{CDAPV09, Oreshkov2012,Oreshkov2012,Branciard2015} and applications \cite{ABCFGB15,Ried2015,CBN2015,Maclean2016quantum,Araujo2014,Guerin2016,Rossi2017}.
However, some important causal structures have not yet received enough attention in the quantum regime. This is the case of the instrumental one.

Partly responsible for that may be the fact \cite{Henson2014} that equipping the common cause with entanglement is not enough to violate the usual \emph{instrumental inequalities} \cite{Pearl1995}.
Instrumental inequalities are to instrumental models what Bell inequalities \cite{Bell1964} are to local hidden-variable ones;
with the difference that instrumental models are intrinsically nonlocal, involving 1-way outcome communication. 
In this sense, instrumental-inequality violations certify a stronger form of nonlocality than Bell violations \cite{Pearl2009}. 
Remarkably, in spite of the no-go result of \cite{Henson2014}, a different class \cite{Bonet2001instrumentality} of instrumental inequalities has been recently shown \cite{Chaves2017} to admit a quantum violation. Besides their fundamental relevance, the violation of instrumental inequalities with quantum resources is also potentially interesting from an applied viewpoint, as it opens a possibility towards new types of nonlocality-based protocols without the requirement of space-like separation, a major experimental overhead to current implementations.

In turn, both instrumental \cite{Pearl1995,Bonet2001instrumentality,Pearl2009,Chaves2017} and Bell \cite{Bell1964,Brunner2014} inequalities are formulated in the device-independent (DI) scenario  of untrusted measurement devices, effectively treated as black boxes with classical settings (inputs) and outcomes (outputs). The DI regime is known to be experimentally much more demanding  \cite{BCWSW12} than the so-called 1-sided (1S) DI one, where one of the observable nodes is a black box while the other one a trusted apparatus with full quantum control.
This is the natural framework of \emph{steering} \cite{RDBCL09}, a hybrid form of quantum nonlocality intermediate between Bell and entanglement. While, in the DI scenario, local hidden-variable models enhanced with different types of communication have a long history in the literature \cite{Maudlin1992, Gisin1999, BCT1999,  Steiner2000, TB03,DegorreLR2005,RegevT09,Pawlowski2010,Chaves2015b,Ringbauer2016,BC17,Chaves2017}, in the 1S-DI setting only input communication has received some attention \cite{SABGS16, NV16}. In contrast, output-communication enhanced local models are totally unexplored in the 1S-DI domain.   

Here, we study \emph{1S quantum instrumental (1SQI) processes} obtained from quantizing the communication-receiving node in classical instrumental causal networks, or, equivalently, from enhancing local hidden-state models with outcome communication.
We introduce \emph{1S-DI instrumental inequalities} and \emph{non-instrumentality witnesses}, as experimentally-friendly tools to test against 1SQI models. These naturally lead to a resource-theoretic measure efficiently computable via semi-definite programming: the \emph{robustness of non-instrumentality}. 
Furthermore, we show that 1S-DI instrumental inequalities can be violated with little entanglement and purity at the common-cause node and, remarkably, without outcome communication, so that the violations are due solely to the states' steering. We present an experimental demonstration in an entangled-photon platform.
Finally, we prove an even stronger incompatibility between steering and 1SQI processes. Namely, that quantum steering alone is enough to attain any value of the robustness. 
Our findings imply that steering is a stronger quantum phenomenon than previously thought, beyond classical hidden-variable models even when equipped with output communication.

\begin{figure}[t!]
\centering
\includegraphics[width=\linewidth]{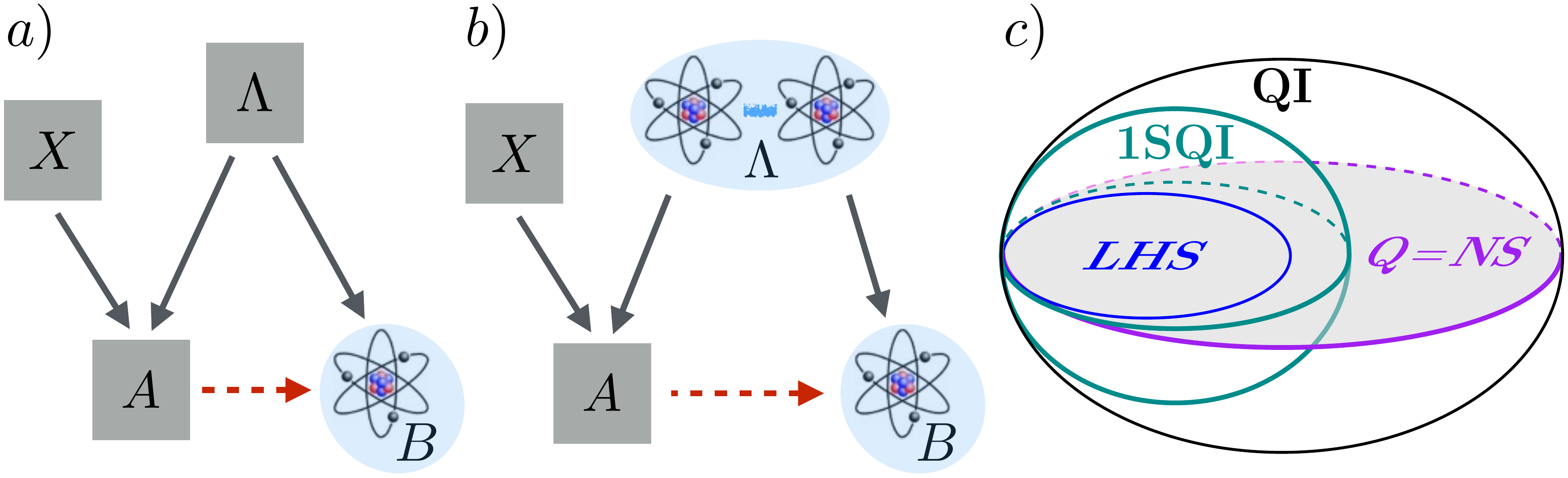}
\caption{$a)$ and $b)$ Hybrid (classical-quantum) instrumental processes. Causal structures are specified by directed acyclic graphs (DAGs). Each node encodes either a classical random variable or a quantum system, and each directed edge a causal influence. The term acyclic refers to the physical requirement of no causal loops. 
$a)$ 
1-sided quantum instrumental ($\mathsf{1SQI}$) causal networks generalise the classical instrumental causal structure to the case where node $B$ is a quantum system.
Removing the red dashed edge, in turn, leads to local hidden-state ($\mathsf{LHS}$) models. $b)$ Quantising also $\Lambda$ gives what we here refer to as quantum instrumental ($\mathsf{QI}$) processes: $\Lambda$ is now a bipartite system in a possibly entangled quantum state, with one subsystem causally influencing $A$ and the other one $B$. Removing the red dashed edge here leads to quantum ($\mathsf{Q}$) models. 
$c)$ Pictorial representation of the inner geometry of the set $\mathsf{QI}$. $\mathsf{1SQI}$ is a strict subset of $\mathsf{QI}$. The lower dimensional manifold $\mathsf{NS}$ of non-signalling assemblages coincides (for the bipartite case) with $\mathsf{Q}$, in turn containing $\mathsf{LHS}$. $\mathsf{LHS}$ and $\mathsf{Q}$ are the sets studied in steering theory. Surprisingly, $\mathsf{Q}$ is not contained in $\mathsf{1SQI}$. What is more, for any assemblage in  $\mathsf{QI}$ and outside of $\mathsf{1SQI}$, there is an assemblage in $\mathsf{Q}$ as far away from $\mathsf{1SQI}$ as the former. 
}
\label{fig:inst}
\end{figure}

\emph{1-sided quantum and quantum-common-cause instrumental processes}. 
We start by introducing 1SQI causal networks, shown in Fig. \ref{fig:inst} a). Nodes $X$, $A$, and $B$ are observable, while $\Lambda$ is hidden. Node $B$ is a quantum system with Hilbert space $\mathbb{H}_{B}$, whereas all four other nodes encode classical random variables. Node $X$ takes  $|X|$ possible values $x\in\mathbb{Z}_X$, with the short-hand notation $\mathbb{Z}_X\coloneqq \{0,\hdots |X|-1\}$ introduced,  $A$  takes $|A|$ values $a\in\mathbb{Z}_A$, and $\Lambda$ can -- w.l.o.g. -- be assumed to take $|\Lambda|=|A|^{|X|}$ possible values $\lambda\in\mathbb{Z}_\Lambda$ (see App. \ref{sec:SDP}). A 1SQI causal model assigns a probability distribution $\boldsymbol{P}_{\Lambda}\coloneqq\{P_{\lambda}\}_{\lambda}$ to $\Lambda$, a  conditional distribution $\boldsymbol{P}_{A\vert X, \Lambda}\coloneqq\{P_{a\vert x, \lambda}\}_{a, x, \lambda}$ to $A$, and a quantum state $\varrho_{a,\lambda}\in\mathcal{B}(\mathbb{H}_{B})$ to $B$, i.e. such that $\varrho_{a,\lambda}\geq 0$ and $\Tr[\varrho_{a,\lambda}]=1$ for all $a$ and $\lambda$. We are interested in the statistics of $A$ and $B$ given $X$. Hence, the local statistics of $X$ is not explicitly considered here. We refer to the users at nodes $A$ and $B$ as Alice and Bob, respectively.

Since $A$ is classical and $B$ quantum, they are most-conveniently described jointly by subnormalised conditional states $\sigma_{a\vert x}$, which encapsulate both the probability $P_{a\vert x}\coloneqq\Tr\left[\sigma_{a\vert x}\right]$ of $a$ given $x$ for Alice and the conditional state $\varrho_{a,x}\coloneqq \sigma_{a\vert x}/P_{a\vert x}$ given $a$ and $x$ for Bob. 1SQI models produce ensembles $\boldsymbol{\Sigma}^{(\rm{inst})}_{A\vert X}\coloneqq\big\{\sigma^{(\rm{inst})}_{a\vert x}\in\mathcal{B}(\mathbb{H}_{B})\big\}_{a, x}$, with
\begin{equation}
\label{eq:assemblage_inst_proc}
\sigma^{(\rm{inst})}_{a\vert x}=\sum_{\lambda}P_{\lambda}\, P_{a\vert x, \lambda}\,\varrho_{a,\lambda}.
\end{equation}
We refer to $\boldsymbol{\Sigma}^{(\rm{inst})}_{A\vert X}$ as a \emph{1SQI assemblage}, and denote the set of all such assemblages by $\mathsf{1SQI}$. 
The term ``assemblage" is native of the steering literature \cite{Cavalcanti2016}. Its use here is not coincidental: there is a connection between $\mathsf{1SQI}$ and steering. To see this, let us next introduce local hidden-state (LHS) models. These correspond to restricted 1SQI models without the causal influence from $A$ to $B$. An \emph{LHS assemblage} $\boldsymbol{\Sigma}^{(\rm{lhs})}_{A\vert X}\coloneqq\big\{\sigma^{(\rm{lhs})}_{a\vert x}\in\mathcal{B}(\mathbb{H}_{B})\big\}_{a, x}$ has components 
$\sigma^{(\rm{lhs})}_{a\vert x}\coloneqq\sum_{\lambda}P_{\lambda}\, P_{a\vert x, \lambda}\,\varrho_{\lambda}$,
i.e. as Eq. \eqref{eq:assemblage_inst_proc} but with  $\varrho_{a,\lambda}$ independent of $a$.
We denote the set of all LHS assemblages by $\mathsf{LHS}$, and call any assemblage $\boldsymbol{\Sigma}_{A\vert X}$ \emph{steerable} if $\boldsymbol{\Sigma}_{A\vert X}\not\in\mathsf{LHS}$.  Clearly, $\mathsf{LHS}\subseteq\mathsf{1SQI}$.

In turn, allowing not only for a quantum $B$ but also for a quantum $\Lambda$ defines the set $\mathsf{QI}$ of \emph{quantum instrumental (QI) assemblages} $\boldsymbol{\Sigma}^{(\rm{qinst})}_{A\vert X}\coloneqq\big\{\sigma^{(\rm{qinst})}_{a\vert x}\in\mathcal{B}(\mathbb{H}_{B})\big\}_{a, x}$. More precisely, we allocate to $\Lambda$ a composite Hilbert space $\mathbb{H}_{\Lambda}=\mathbb{H}_{\Lambda_A}\otimes\mathbb{H}_{\Lambda_B}$, such that subsystems $\Lambda_A$ and $\Lambda_B$ causally influence nodes $A$ and $B$, respectively [see Fig. \ref{fig:inst} b)]. Hence, $\Lambda$ is now a \emph{quantum common cause} \cite{Costa2016, Allen2016} for $A$ and $B$ \cite{ThirdComment}.
Accordingly, for $\Lambda$ in a state $\varrho_{\Lambda}\in\mathcal{B}(\mathbb{H}_{\Lambda})$, the resulting QI conditional states are 
\begin{align}
\label{eq:assemblage_qi}
\sigma^{(\rm{qinst})}_{a\vert x}&\coloneqq\mathcal{E}_{B\vert a,\Lambda_B}\left(\Tr_{\Lambda_A}\left[M^{(a)}_x\otimes \openone_{\Lambda_B}\,\varrho_{\Lambda}\right]\right).
\end{align}
Here,  $\mathcal{E}_{B\vert a,\Lambda_B}:\mathcal{B}(\mathbb{H}_{\Lambda_B})\to\mathcal{B}(\mathbb{H}_{B})$ is an $a$-dependent completely-positive trace-preserving  map and 
 $M^{(a)}_x$ is the $a$-th element of an $x$-dependent measurement $\boldsymbol{M}_x\coloneqq\big\{M^{(a)}_x\in\mathcal{B}(\mathbb{H}_{\Lambda_A}):M^{(a)}_x\geq 0,\ \sum_{a'}M^{(a')}_x=\openone_{\Lambda_A}\big\}_{a}$, with $\openone_{\Lambda_A}$ the identity on $\mathbb{H}_{\Lambda_A}$. Clearly, $\mathsf{1SQI}\subseteq\mathsf{QI}$, as Eq. \eqref{eq:assemblage_qi} reduces to Eq. \eqref{eq:assemblage_inst_proc} for the specific case of $\varrho_{\Lambda}$ separable.

On the other hand, for the particular case of $\mathcal{E}_{B\vert a,\Lambda_B}$ being the identity map for all $a$ (no causal influence from $A$ to $B$), $\mathsf{QI}$ reduces to the set $\mathsf{Q}$ of \emph{quantum assemblages} $\boldsymbol{\Sigma}^{(\rm{q})}_{A\vert X}\coloneqq\big\{\sigma^{(\rm{q})}_{a\vert x}\in\mathcal{B}(\mathbb{H}_{B})\big\}_{a, x}$, of components
$\sigma^{(\rm{q})}_{a\vert x}\coloneqq\Tr_{\Lambda_A}\big[M^{(a)}_x\otimes \openone_{\Lambda_B}\,\varrho_{\Lambda}\big]$.
Hence, $\mathsf{Q}$ is to $\mathsf{QI}$ what $\mathsf{LHS}$ is to $\mathsf{1SQI}$. Clearly, $\mathsf{Q}\subseteq\mathsf{QI}$. In addition, from steering theory, we know that $\mathsf{LHS}\subset\mathsf{Q}$.
On the contrary, it holds that $\mathsf{1SQI}\not\subset\mathsf{Q}$, as Q assemblages are non-signalling while 1SQI ones not. An assemblage $\boldsymbol{\Sigma}_{A\vert X}$ is said to be non-signalling if 
\begin{equation}
\label{eq:lhs_NS}
\varrho_{B}\coloneqq\sum_{a}\sigma_{a\vert x}
\end{equation}
(the reduced state of Bob) is independent of $x$. Also due to non-signalling, it follows that, actually, $\mathsf{LHS}\subset\mathsf{1SQI}$ and $\mathsf{Q}\subset\mathsf{QI}$.
We call the set of non-signalling assemblages $\mathsf{NS}$. For the bipartite case under consideration, it is known  that $\mathsf{Q}=\mathsf{NS}$ \cite{SBCSV15}. In contrast, for instrumental causal models, Bob's state can depend on $x$ even after summing $a$ out.

\emph{1S-DI instrumental inequalities, witnesses, and robustness}. Since $\mathsf{1SQI}$ is convex, any  $\boldsymbol{\Sigma}_{A\vert X}\notin\mathsf{1SQI}$ is separated from $\mathsf{1SQI}$ by a hyperplane, represented by an assemblage-like object $\boldsymbol{W}_{A\vert X}\coloneqq\big\{W_{a\vert x}\in\mathcal{B}(\mathbb{H}_{B})\big\}_{a, x}$, with $W_{a\vert x}$ Hermitian, of fixed \emph{scale} $s\coloneqq\sum_{a,\,x} \Tr\left[W_{a\vert x}\right]$, such that 
\begin{equation}
\label{eq:1SQI_inequality}
\left\langle \boldsymbol{W}_{A\vert X}, \boldsymbol{\Sigma}^{(\rm{inst})}_{A\vert X} \right\rangle\coloneqq\sum_{a,\,x} \Tr\left[W_{a\vert x}\,\sigma^{(\rm{inst})}_{a\vert x}\right]\leq\beta,
\end{equation}
for all $\boldsymbol{\Sigma}^{(\rm{inst})}_{A\vert X}\in\mathsf{1SQI}$,
and $\big\langle \boldsymbol{W}_{A\vert X}, \boldsymbol{\Sigma}_{A\vert X} \big\rangle>\beta\in\mathbb{R}$. 
We refer to Eq. \eqref{eq:1SQI_inequality} as a 1S quantum instrumental inequality with \emph{1SQI bound} $\beta$ (which depends solely on $\boldsymbol{W}_{A\vert X}$). 
Thus, $\boldsymbol{W}_{A\vert X}$ plays a role analogous to the normal vector of a plane in Euclidean space. We refer to $\boldsymbol{W}_{A\vert X}$ as a \emph{non-instrumentality witness}. 
The separation is then quantified by the violation $\langle \boldsymbol{W}_{A\vert X}, \boldsymbol{\Sigma}_{A\vert X} \rangle-\beta$. 
Finally, we say that $\boldsymbol{W}_{A\vert X}$ is an \emph{optimal non-instrumentality witness} for $\boldsymbol{\Sigma}_{A\vert X}\notin\mathsf{1SQI}$ if $\langle \boldsymbol{W}_{A\vert X}, \boldsymbol{\Sigma}_{A\vert X} \rangle - \beta\geq\langle \boldsymbol{W}'_{A\vert X}, \boldsymbol{\Sigma}_{A\vert X} \rangle - \beta'$ for all non-instrumentality witnesses $\boldsymbol{W}'_{A\vert X}$ with 1SQI bound $\beta'$ and scale $\sum_{a,\,x} \Tr\big[W'_{a\vert x}\big]=s$. 
Remarkably, as shown in App. \ref{sec:SDP}, the optimal witness is obtained efficiently  via semi-definite programming \cite{Cavalcanti2016}.

\begin{figure}[t!]
\centering
\includegraphics[width=\linewidth]{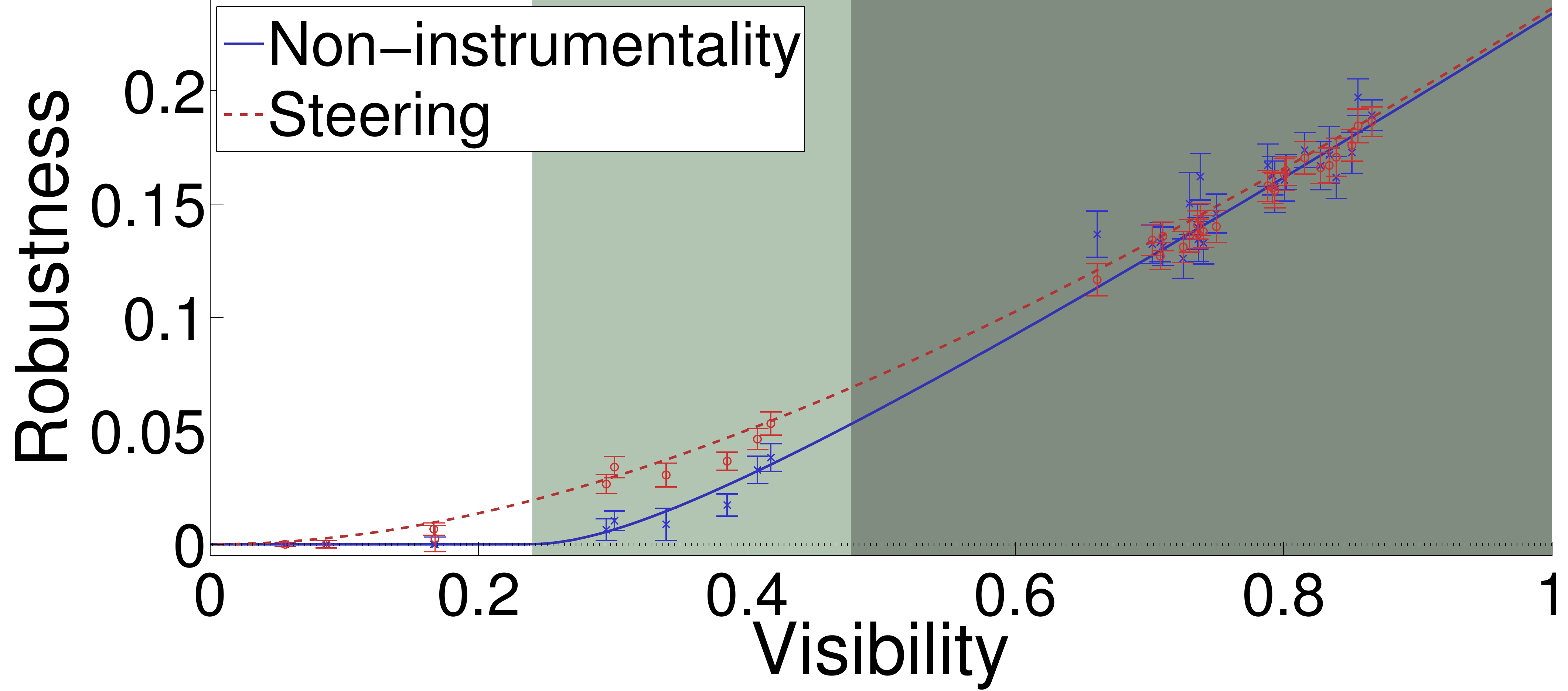}
\caption{Robustnesses of steering and of non-instrumentality. The curves and points correspond respectively to theory and experiment. An assemblage is produced by 3 projective local measurements by Alice on the
$\varrho_\Lambda=V\ket{\Phi^+}\bra{\Phi^+}+\frac{(1-V)}{2}(\ket{00}\bra{00}+\ket{11}\bra{11})$, generated by local dephasing on the maximally entangled state $\ket{\Phi^+}=(1/\sqrt{2})(\ket{00}+\ket{11})$ with strength given by the visibility $V$. 
The assemblage is steerable in the usual sense for all $V>0$, and is compatible with 1SQI models up to $V < 0.24$ (white region).
That point is the onset of a stronger form of  quantum steering  than explainable by 1SQI models (light and dark green). 
Finally, for $V > (9-4\sqrt{2})/7\simeq 0.48$ (dark green), the assemblage produces (under the optimal  \cite{Taddei2016} $a$-dependent measurements on Bob's qubit) black-box correlations that violate the DI instrumental inequality of \cite{Bonet2001instrumentality,Chaves2017} (see App. \ref{sec:DI_Inst_inequality}). That is, the assemblage's incompatibility with $\mathsf{1SQI}$ can be verified in the 1S DI setting for a broader range of visibilities than in the fully DI one. Error bars obtained  assuming Poissonian distributions for the photon counts.
}
\label{fig:Qvs1SQI}
\end{figure}

Witnesses are, in turn, connected  with robustness measures \cite{Vidal1999, PW15}. Here, we consider the \emph{robustness of non-instrumentality} $R_{\rm{ni}}$, defined, for any  $\boldsymbol{\Sigma}_{A\vert X}\in\mathsf{QI}$, as 
\begin{equation}
\label{eq:def_robustness}
R_{\rm{ni}}(\boldsymbol{\Sigma}_{A\vert X})\coloneqq\min\bigg\{t\in\mathbb{R}_{\geq0}: \frac{\boldsymbol{\Sigma}_{A\vert X}+t\,\boldsymbol{\Pi}_{A\vert X}}{1+t}\in\mathsf{1SQI}\bigg\}.
\end{equation}
It measures the minimal mixing $t/(1+t)$ with any $\boldsymbol{\Pi}_{A\vert X}\in\mathsf{1SQI}$ that $\boldsymbol{\Sigma}_{A\vert X}$ tolerates before the mixture enters $\mathsf{1SQI}$. 
Interestingly, $R_{\rm{ni}}$ is a measure of non-instrumentality in the formal, resource-theoretic sense \cite{Gallego2015,Gallego2017, Amaral2017}, as we explicitly show in App. \ref{sec:monotonicity}. Moreover, we note that other choices of ``noise" types are possible, giving rise to different variants of $R_{\rm{ni}}$. However, the choice $\boldsymbol{\Pi}_{A\vert X}\in\mathsf{1SQI}$ is particularly convenient as it yields the  robustness efficiently computable through a semi-definite programming optimisation. In fact, such optimisation shows that $R_{\rm{ni}}(\boldsymbol{\Sigma}_{A\vert X})=\langle \boldsymbol{V}_{A\vert X}, \boldsymbol{\Sigma}_{A\vert X} \rangle-1$, where $\boldsymbol{V}_{A\vert X}$ is the optimal for $\boldsymbol{\Sigma}_{A\vert X}$ over a simple subclass of non-instrumentality witnesses (see App. \ref{sec:Robustness_SDP} for details). We call $\boldsymbol{V}_{A\vert X}$ the \emph{optimal robustness witness} for $\boldsymbol{\Sigma}_{A\vert X}$.

Fig. \ref{fig:Qvs1SQI} shows $R_{\rm{ni}}$, together with the usual steering robustness \cite{Cavalcanti2016}, for Q assemblages obtained from local measurements (by Alice) on states with different degrees of entanglement and purity. Some assemblages in the figure have  positive steering robustness and $R_{\rm{ni}}=0$. This confirms that $\mathsf{LHS}\subset\mathsf{1SQI}$:  Alice's outcome signalling indeed provides the models with more descriptive power. However, the figure also shows assemblages with $R_{\rm{ni}} > 0$. This implies the following theorem, proven also analytically in App. \ref{sec:Proof_Theo1}.
\begin{thm}[$\mathsf{Q}\not\subseteq\mathsf{1SQI}$]
\label{theo:one}
Outcome signalling is not enough for LHS models to reproduce quantum steering.
\end{thm}

It is instructive to compare with the fully DI case, where Bob's instrument is also a black box. 
There, usual Bell-like correlations (where Alice and Bob have independent inputs) obtained without outcome communication are known to be stronger than local hidden-variable models augmented with outcome signalling from Alice \cite{Chaves2015b,Ringbauer2016,Ringbauer2017}.
However, for DI instrumental processes, in contrast, if Bob does not actively exploit Alice's output his measurement setting is fixed \cite{FirstComment}.
Hence, no matter how entangled $\varrho_{\Lambda}$, or what measurements Alice makes, the resulting correlations will be automatically compatible with local hidden-variable models with no input for Bob, a subclass of classical instrumental models. In other words, if Bob applies a measurement that does not depend on $a$, the  correlations trivially fulfill any DI instrumental inequality, including the recent one of Ref. \cite{Chaves2017}. Thus, incompatibility with the instrumental DAG without output signalling is a distinctive feature of the 1-sided DI case. 

Finally, $R_{\rm{ni}}$ could in principle attain higher values over $\mathsf{QI}$ than over $\mathsf{Q}$. After all, the former allows for signalling while the latter does not. Surprisingly, this is false. The following theorem, proven in App. \ref{sec:proof_main_theo}, holds instead.
\begin{thm}[$R_{\rm{ni}}(\mathsf{Q})=R_{\rm{ni}}(\mathsf{QI})$]
\label{theo:two}
For every quantum instrumental assemblage, there exists a Q assemblage with the same non-instrumentality robustness.
\end{thm}
From a practical viewpoint, the theorem provides a significant computational shortcut in the task of, given a fixed value of $R_{\rm{ni}}$, finding an assemblage with that robustness. Because the theorem allows one to restrict the search to quantum assemblages, instead of searching over all QI ones \cite{SecondComment}. From a fundamental perspective, in turn, it has implications in the inner geometry of $\mathsf{QI}$. Namely, it tells us that, for any point $\boldsymbol{\Sigma}^{(\rm{qinst})}_{A\vert X}\in\mathsf{QI}\setminus\mathsf{1SQI}$, there is always a point $\boldsymbol{\Sigma}^{(\rm{q})}_{A\vert X}\in\mathsf{Q}$ as far away from $\mathsf{1SQI}$ as $\boldsymbol{\Sigma}^{(\rm{qinst})}_{A\vert X}$. This does not contradict the fact that $\mathsf{Q}\subset\mathsf{QI}$, because $\mathsf{Q}$ is a lower-dimensional manifold (the NS one) of $\mathsf{QI}$. Theorem \ref{theo:two} thus suggests that $\boldsymbol{\Sigma}^{(\rm{q})}_{A\vert X}$ is a kind of projection of $\boldsymbol{\Sigma}^{(\rm{qinst})}_{A\vert X}$ onto $\mathsf{NS}$ [see Fig. \ref{fig:inst} c)].

\begin{figure}[t!]
\centering
\includegraphics[width=\linewidth]{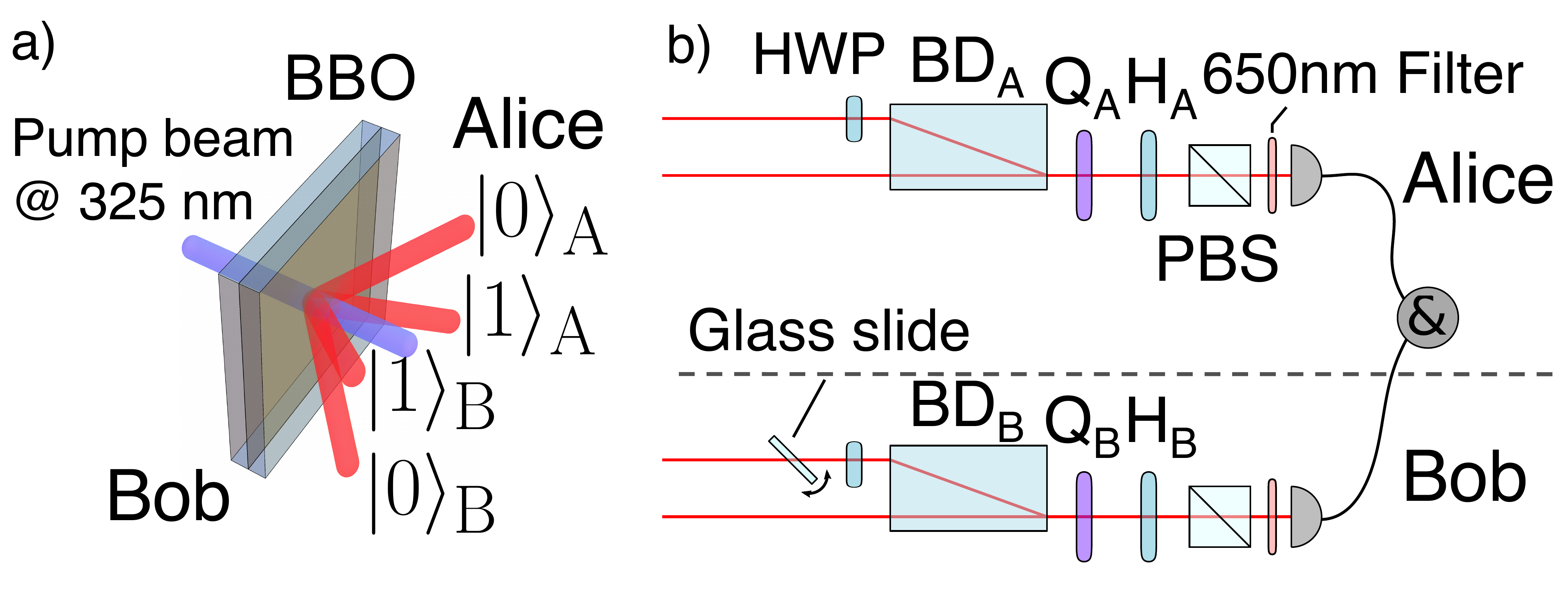}
\caption{Experimental setup. (a) The BBO crystal is pumped by a vertically polarized laser beam at $325$ nm from a He-Cd source, so that a pair of down-converted photons, at $650$ nm, emerge in an entangled state of momenta due to momentum conservation. The momenta encode the two common-cause qubits. All but two momentum modes, corresponding to $|00\rangle$ and $|11\rangle$, are filtered out. The produced momentum state is the two-qubit maximally entangled state $\ketbra{\Phi^+}{\Phi^+}$. One  qubit is sent to Alice and the other one to Bob.
(b) Side view of each party's device, depicting the preparation (by Alice) and measurement (by Bob) of the assemblage. Beam displacers $BD_A$ and $BD_B$ and half-wave plates 
(HWP) map the momenta of each photon into its polarisations. A glass slide introduces dephasing between Bob's two paths, allowing us to tune the visibility $V$ of the resulting dephased state $\varrho_{\Lambda}$  from $V\sim0$ to $V\sim94,5\%$. Quarter-wave plates ($Q_A$ and $Q_B$), half-wave plates ($H_A$ and $H_B$), polarised beam splitters (PBS), and a coincidence-photon detector implement projective polarization measurements. 
}
\label{fig:Diagram}
\end{figure}

\emph{Experimental demonstration}. We verified the gap between quantum steering and non-instrumentality using entangled-photon pairs produced by spontaneous parametric down conversions in a BBO crystal \cite{Farias2012, Aguilar2014, Cavalcanti2015} (see Fig. \ref{fig:Diagram}). 
Alice's device is taken as the black box, so her measurement settings (wave-plate angles) and outcomes encode the bits $x$ and $a$, respectively. Each measurement by her accounts for the preparation of the assemblage element $\sigma^{(\rm{q})}_{a\vert x}$. 
In contrast, Bob's measurements allow for state tomography of each $\sigma^{(\rm{q})}_{a\vert x}$. We produce assemblages of different purities and steering by applying local dephasing of different strengths $V$ on Bob's qubit and having Alice measure the observables $-(\sigma_x + \sigma_z)/\sqrt{2}$, $\sigma_x$, or $\sigma_z$, where  $\sigma_x$ and $\sigma_z$ are respectively the first and third Pauli matrices. These three observables correspond to the input choices $x=$0, 1, and $2$, respectively. We tomographically reconstruct each produced assemblage and evaluate $R_{\rm{ni}}$, whose values are those displayed in Fig. \ref{fig:Qvs1SQI}.

\emph{Final discussion}. 1SQI processes are a generalisation of both classical instrumental processes (native of causal inference \cite{Spirtes2001,Pearl2009}) to the case where the final node is quantum and of local hidden-state models (native of steering theory \cite{RDBCL09,Cavalcanti2016}), thus unifying two previously disconnected topics. We introduced inequalities and witnesses to test against such 1SQI causal models. These can, in particular, detect entanglement with a single trusted device even in the presence of 1-way outcome signalling. Strikingly, they can also be violated with entanglement alone, i.e. without using outcome communication. Hence, 
outcome communication is not enough to explain quantum steering.
Interestingly, this is a distinctive feature of the 1S DI setting: For DI instrumental processes, if Bob's setting is independent of Alice's outcome, any state produces local hidden-variable correlations, automatically compatible with classical instrumental models.

The proposed quantifier (robustness) of non-instrumentality is efficiently computable via semi-definite programming. Moreover, we prove in the appendix that it is a formal, resource-theoretic monotone
.
With it, we showed an even stronger incompatibility between steering and 1SQI processes: that any value of the robustness can be attained with steering alone. We experimentally verified our predictions in an entangled-photon platform. The experiment is simple, but proves that quantum states can be steerable in a stronger way than previously reported
.

Finally, the fact that quantum mechanics allows one to falsify -- with  quantum control only at a single lab -- classical explanations even when these exploit output signalling is not only relevant from the perspectives of quantum foundations and causal inference but also promising from an applied one. 
More precisely, quantum-nonlocality applications are experimentally less demanding in the 1S DI regime than the fully DI one, as already mentioned. To this, our findings now add the possibility of steering-based protocols with the additional experimentally-appealing feature of no need for space-like separation. 
We note that, even if $B$ is in the future light-cone of $A$ (and, therefore, also of $X$), direct causal influences from $X$ to $B$ can be ruled out (and so an underlying instrumental causal structure guaranteed) with interventions on $A$, e.g. There, cryptographic or randomess-generation protocols based on 1S quantum instrumental inequality violations are conceivable.
It would thus be interesting to explore steering beyond outcome signalling as a potential resource for information processing in comparison with conventional steering-based schemes \cite{BCWSW12} requiring space-like separation.  
Our results open new venues for research in that direction. \nocite{ProgramsLink}


\begin{acknowledgments}
We are indebted to Stephen P. Walborn for the experimental infrastructure and thank Ana B. Sainz for noticing an error in a previous version. LA and RC acknowledge support from the Brazilian ministries MCTIC and MEC. LA, MMT, and RVN acknowledge support from the Brazilian agencies CNPq, FAPERJ, and INCT-IQ; and LA also from FAPESP.
\end{acknowledgments}


%


\newpage
\appendix

\section{1-sided quantum instrumentality as a semi-definite programming membership problem}
\label{sec:SDP}
In this section, we consider the problems of how to determine if a given arbitrary assemblage $\boldsymbol{\Sigma}_{A\vert X}$ is or not in $\mathsf{1SQI}$ and how to determine its optimal non-instrumentality witness, together with its corresponding violation. As in the membership problem for $\mathsf{LHS}$ and the determination of optimal steering witnesses of standard steering theory \cite{Cavalcanti2016}, these problems turn out to admit a formulation as a semi-definite programe (SDP). SDPs deal with optimisations of a linear objective function over a matrix space defined by linear and positive-semidefinite constraints. Because of this, SDPs are exact in the sense that the solutions they return are guaranteed not to get stuck at local maxima or minima \cite{boyd2004convex}.

To this end, we express the conditional states in Eq. \eqref{eq:assemblage_inst_proc} as
\begin{equation}
\sigma^{(\rm{inst})}_{a\vert x}=\sum_{\tilde{\lambda}} D_{\tilde{\lambda}}(a\vert x)\,\tilde{\sigma}_{a,\tilde{\lambda}},
\label{eq:Rho_QIDAG_Det}
\end{equation}
where $D_{\tilde{\lambda}}$ is the $\tilde{\lambda}$-th deterministic response function and $\tilde{\sigma}_{a,\tilde{\lambda}}\coloneqq\sum_{\lambda}\tilde{P}_{\tilde{\lambda}\vert\lambda}\,P_{\lambda}\, \varrho_{a,\lambda}$, with $\tilde{P}_{\tilde{\lambda}\vert\lambda}$ defined such that $P_{a\vert x, \lambda}\eqqcolon\sum_{\tilde{\lambda}}\tilde{P}_{\tilde{\lambda}\vert\lambda}\,D_{\tilde{\lambda}}(a\vert x)$. There are as many such functions as hidden-variable values, i.e. $|\tilde{\Lambda}|=|\Lambda|=|A|^{|X|}$. In addition, the conditional states $\tilde{\sigma}_{a,\tilde{\lambda}}$ are subnormalised such that 
\begin{equation}
\Tr\left[\tilde{\sigma}_{a,\tilde{\lambda}}\right]=\sum_{\lambda} \tilde{P}_{\tilde{\lambda}\vert\lambda}\,P_{\lambda}\eqqcolon \tilde{P}_{\tilde{\lambda}},
\label{eq:independ_a}
\end{equation} for all $a\in\mathbb{Z}_A$ and all $\tilde{\lambda}\in\mathbb{Z}_{\Lambda}$, i.e. their trace is independent of $a$. Note that the distribution $\boldsymbol{\tilde{P}}_{\Lambda}\coloneqq\big\{\tilde{P}_{\tilde{\lambda}}\big\}_{\tilde{\lambda}}$ is automatically normalized if so is $\boldsymbol{\Sigma}_{A\vert X}$.

We can then recast the membership problem of $\boldsymbol{\Sigma}_{A\vert X}$ for $\mathsf{1SQI}$, i.e. whether $\boldsymbol{\Sigma}_{A\vert X}$ admits or not a decomposition as in Eq. \eqref{eq:assemblage_inst_proc}, directly as an SDP feasibility test. This can be conveniently expressed by the following optimisation.
\begin{subequations}
\label{eq:primal_given}
\begin{align}
\nonumber
\textrm{Given} &\quad \boldsymbol{\Sigma}_{A\vert X}=\big\{\sigma_{a\vert x}\big\}_{a, x},  \\
\label{eq:SDP_StrictFeasible_objective}
\min_{
\{ \tilde{\sigma}_{a,\lambda}\in\mathcal{B}(\mathbb{H}_{B})\}_{a, \lambda}}  &\quad \mu, \\
\label{eq:SDP_StrictFeasible_decomposition}
\textrm{s. t.} &\quad \sigma_{a\vert x}=\sum_{\lambda} D_{\lambda}(a\vert x)\,\tilde{\sigma}_{a,\lambda}, \\
\label{eq:SDP_StrictFeasible_probabilities}
\textrm{with} &\quad \Tr\left[ \tilde{\sigma}_{a,\lambda}\right]=\,\Tr\left[\tilde{\sigma}_{0,\lambda}\right], \\
\label{eq:SDP_StrictFeasible_Primal}
\textrm{and} &\quad \tilde{\sigma}_{a,\lambda} \geq -\mu\,\openone_B.
\end{align}
\end{subequations}
Eqs. \eqref{eq:SDP_StrictFeasible_decomposition} and \eqref{eq:SDP_StrictFeasible_probabilities} encode the constraints in Eqs. \eqref{eq:Rho_QIDAG_Det} and \eqref{eq:independ_a}, respectively. Hence, the minimisation in Eq. \eqref{eq:SDP_StrictFeasible_objective} amounts to finding an 1SQI decomposition in terms of conditional states $\tilde{\sigma}_{a,\tilde{\lambda}}$ as positive as possible, in the sense of satisfying the constraint of Eq. \eqref{eq:SDP_StrictFeasible_Primal} with $\mu$ as negative as possible. When the objective function $\mu$ reaches a non-positive value, a decomposition as in Eq. \eqref{eq:Rho_QIDAG_Det} is feasible with some $\tilde{\sigma}_{a,\tilde{\lambda}}\geq0$, and vice versa. That is, any value $\mu>0$ returned by the optimisation is equivalent to an 1SQI-decomposition being infeasible for $\boldsymbol{\Sigma}_{A\vert X}$, i.e. to $\boldsymbol{\Sigma}_{A\vert X}\not\in\mathsf{1SQI}$. 

By virtue of the duality theory of semi-definite programming \cite{Cavalcanti2016,boyd2004convex}, every such SDP admits a dual, equivalent formulation as follows.
\begin{subequations}
\label{eq:dual_given}
\begin{align}
\nonumber
\textrm{Given} &\quad \boldsymbol{\Sigma}_{A\vert X}=\big\{\sigma_{a\vert x}\big\}_{a, x}, \nonumber \\
\label{eq:dualstrict_min_W}
\max_{\{W_{a\vert x}\in\mathcal{B}(\mathbb{H}_{B})\}_{a, x}}  &\quad\sum_{a,\, x} \Tr\left[W_{a\vert x}\,\sigma_{a\vert x}\right],\\
\label{eq:first_constraint}
\textrm{s. t.} &\quad\sum_{x}\,W_{a\vert x}\,D_\lambda(a\vert x)\,\leq C_{a,\lambda}\,\openone_B,  \\
\label{eq:second_constraint}
\textrm{with} &\quad \sum_{a} C_{a,\lambda}=0,\\
\label{eq:dualstrict_scale_constraint}
\textrm{and} &\quad \sum_{a,\, x,\, \lambda}\,\Tr\left[W_{a\vert x}\,D_\lambda(a\vert x)\right]=-1.
\end{align}
\end{subequations}
where $C_{a,\lambda}\in\mathbb{R}$ for all $a\in\mathbb{Z}_{A}$ and all $\lambda\in\mathbb{Z}_{\Lambda}$. 
Eqs. \eqref{eq:first_constraint} and \eqref{eq:second_constraint} together imply that $\sum_{a,\, x}\Tr\left[W_{a\vert x}\, \sum_{\tilde{\lambda}}D_{\tilde\lambda}(a\vert x)\,\tilde{\sigma}_{a,\tilde{\lambda}}\right]\,\leq0$, for any conditional states $\tilde{\sigma}_{a,\tilde{\lambda}}$ satisfying Eq. \eqref{eq:independ_a}. So, these two equations encode the constraints that the assemblage-like object $\boldsymbol{W}_{A\vert X}=\{W_{a\vert x}\in\mathcal{B}(\mathbb{H}_{B})\}_{a, x}$ of Hermitian operators $W_{a\vert x}$ returned by the optimisation is a non-instrumentality witness for some $\beta\geq0$. 
Eq. \eqref{eq:dualstrict_scale_constraint} fixes the scale of $\boldsymbol{W}_{A\vert X}$, which prevents the maximisation in Eq.  \eqref{eq:dualstrict_min_W} from diverging to $\infty$. Indeed, using the fact that $\sum_{\lambda}\,D_\lambda(a\vert x)=|A|^{|X|-1}$, Eq.  \eqref{eq:dualstrict_scale_constraint} yields 
\begin{equation}
\label{eq:scaling_W}
s=-\frac{1}{|A|^{|X|-1}}.
\end{equation}
Other choices of scaling are valid, but they must be accompanied by a corresponding rescaling factor for the primal objective function in Eq. \eqref{eq:SDP_StrictFeasible_objective}. 
Finally, the maximisation in Eq. \eqref{eq:dualstrict_min_W} guarantees that  $\boldsymbol{W}_{A\vert X}$ is the  optimal non-instrumentality witness for $\boldsymbol{\Sigma}_{A\vert X}$ and that it is therefore (due to the convexity of $\mathsf{1SQI}$) tight -- i.e. that $\beta=0$ --, as we wanted to show.
In other words, the maximisation returns a positive value if, and only if, $\boldsymbol{\Sigma}_{A\vert X}\notin\mathsf{1SQI}$. 

Finally, it is important to mention that the primal and dual SDPs, given respectively by Eqs. \eqref{eq:primal_given} and \eqref{eq:dual_given}, satisfy a convenient property called \emph{strong duality} \cite{Cavalcanti2016,boyd2004convex}. By virtue of this, the primal and dual objective functions, i.e. $\mu$ and $\sum_{a,\, x} \Tr\left[W_{a\vert x}\,\sigma_{a\vert x}\right]$, respectively, converge to the same optimal values. That is, the minimum of Eq. \eqref{eq:SDP_StrictFeasible_objective} and the maximum of Eq. \eqref{eq:dualstrict_min_W} are guaranteed to coincide.

\section{Proof of theorem 1}
\label{sec:Proof_Theo1}
Here, we analytically prove theorem 1. We give a non-instrumentality witness $\boldsymbol{W}_{A|X}$ and a quantum assemblage $\boldsymbol{\Sigma}^{(\rm{q})}_{A\vert X}\in\mathsf{Q}$ such that $\boldsymbol{\Sigma}^{(\rm{q})}_{A\vert X}$ violates the 1S quantum instrumental inequality defined by $\boldsymbol{W}_{A|X}$. The assemblage we use is the same as that of Fig. 2 in the main text for $V=1$ (no dephasing), i.e. the one obtained from $|\Phi^+\rangle = \left(|00\rangle + |11\rangle\right)/\sqrt{2}$ through projective local measurements by Alice in the bases $-(\sigma_x + \sigma_z)/\sqrt{2}$, $\sigma_x$, and $\sigma_z$. As our witness, we use the assemblage itself multiplied by a factor 2 for normalization purposes, i.e. $\boldsymbol{W}_{A|X}= 2\,\boldsymbol{\Sigma}^{(\rm{q})}_{A\vert X}$. Below, we show that the 1SQI bound corresponding to this witness is $\beta=2 + 1/\sqrt{2}$. On the other hand, it is immediate to see that
\begin{equation}
 \big\langle\boldsymbol{W}_{A|X},\,\boldsymbol{\Sigma}^{(\rm{q})}_{A|X}\big\rangle = 3>2 + 1/\sqrt{2}.
 \label{eq:AnProof_WitnessAssemblage} 
\end{equation}
 This implies that $\boldsymbol{\Sigma}^{(\rm{q})}_{A\vert X}\notin\mathsf{1SQI}$ and, therefore, that $\mathsf{Q}\not\subseteq\mathsf{1SQI}$.

To prove that $\beta=2 + 1/\sqrt{2}$, we analytically maximise $\big\langle \boldsymbol{W}_{A|X},\,\boldsymbol{\Sigma}^{(\rm{inst})}_{A|X}\big\rangle$ over all $\boldsymbol{\Sigma}^{(\rm{inst})}_{A|X}\in\mathsf{1SQI}$ and obtain the claimed maximum $2 + 1/\sqrt{2}$.
To this end, note first that the components of the quantum assemblage $\boldsymbol{\Sigma}^{(\rm{q})}_{A|X}$ in question are rank 1, so that $\Tr\left[ W_{a,x}\, W_{a^\prime,x}\right] = \delta_{a,a^\prime}$. Then, using that $\sigma^{(\rm{inst})}_{a|x} = \sum_{\lambda \in \Lambda} P(\lambda)\,D_\lambda(a|x)\,\rho_{a,\,\lambda}$, one gets
\begin{equation}
\big\langle\boldsymbol{W}_{A|X},\,\boldsymbol{\Sigma}^{(\rm{inst})}_{A|X}\big\rangle = \frac{3}{2} + \frac{1}{2}P(\lambda)\left(\boldsymbol{v}_{0,\lambda}\cdot\boldsymbol{r}_{0,\lambda} - \boldsymbol{v}_{1,\lambda}\cdot\boldsymbol{r}_{1,\lambda}\right),
\label{eq:AnProof_WitnessSQI}
\end{equation}
where $\boldsymbol{r}_{a,\lambda} \coloneqq~\sum_{j=1}^{3} \Tr[\rho_{a,\lambda}\,\sigma_j]\,\boldsymbol{e}_j$ and
\begin{equation}
\boldsymbol{v}_{a,\lambda} \coloneqq \sum_{x = 0}^{2} D_\lambda(a|x)\,\boldsymbol{u}_{x},
\end{equation}
with $\boldsymbol{u}_{x} \coloneqq~\sum_{j=1}^{3} \Tr[W_{0,x}\,\sigma_j]\,\boldsymbol{e}_j$ and $\boldsymbol{e}_j$ the $j$-th element of the canonical orthogonal basis of $\mathbb{R}^3$, for $j=1$, 2, or 3. Finally, optimising over $P(\lambda)$ and $\boldsymbol{r}_{a,\lambda}$ yields
\begin{equation}
\big\langle\boldsymbol{W}_{A|X},\,\boldsymbol{\Sigma}^{(\rm{inst})}_{A|X}\big\rangle= \frac{3}{2} + \frac{1}{2}\max_{\lambda \in \Lambda} \| \boldsymbol{v}_{0,\lambda} \| + \| \boldsymbol{v}_{1,\lambda}\|.
\label{eq:AnProof_SQIThreshold}
\end{equation}
For the vectors $\boldsymbol{u}_x$ used here (namely $\boldsymbol{u}_0 = -(\boldsymbol{e}_1 + \boldsymbol{e}_3)/\sqrt{2}$, $\boldsymbol{u}_1 = \boldsymbol{e}_1$, and $\boldsymbol{u}_2 = \boldsymbol{e}_3$) we obtain $\max_{\lambda \in \Lambda} \| \boldsymbol{v}_{0,\lambda} \| + \| \boldsymbol{v}_{1,\lambda}\| = 1 + \sqrt{2}$, from which the value $\beta=~2 + 1/\sqrt{2}$ follows and the proof is completed.

\section{Device-independent instrumental inequality}
\label{sec:DI_Inst_inequality}
We use the linear inequality derived in \cite{Bonet2001instrumentality} and recently revisited in \cite{Chaves2017} to test for violations of the classical instrumental model by our assemblages, under an $a$-dependent measurement $M_a$ by Bob (his input is equal to the output of Alice's black box). The inequality can be expressed as
\begin{align}
\nonumber
\sum_{a,\ b \in \{0,1\}} \left[(-1)^a - (-1)^b - (-1)^{a+b}\right]\,P(a,b|0)&  \\
+ 2\,(-1)^b\,P(a,b|1) + 2 (-1)^{a+b} \,P(a,b|2)&\leq 3,
\label{DI_ineq}
\end{align}
where $P(a,b|x) \coloneqq \Tr\left[M^{(b)}_a \, \sigma_{a|x}\right]$, with $M^{(b)}_a$ the $b$-th element of the $a$-th measurement of Bob's. The optimal measurements by Bob for the maximal violation are  obtained through the  analytical technique of Ref. \cite{Taddei2016}.

\section{Monotonicity of $R_{\rm{ni}}$ under free operations of non-instrumentality}
\label{sec:monotonicity}
In this section we prove that $R_{\rm{ni}}$ is a \emph{non-instrumentality monotone} for any linear class of free operations of non-instrumentality. We leave the details of the resource theory of non-instrumentality (in particular the explicit form of the corresponding free operations) for future work and prove monotonicity solely from the abstract generic properties of free operations. That is, we prove that $R_{\rm{ni}}$ is monotonous (non-increasing) under any linear map satisfying the essential free-operation requirement that $\mathcal{M}(\boldsymbol{\Sigma}_{A\vert X})\in\mathsf{1SQI}$ for all $\boldsymbol{\Sigma}_{A\vert X}\in\mathsf{1SQI}$.  The proof is similar to that \cite{Gallego2015} of steering monotonicity for the steering robustness \cite{Cavalcanti2016}.

By definition, $R_{\rm{ni}}(\boldsymbol{\Sigma}_{A\vert X})$ is the minimal value $t^*\in \mathbb{R}_{\geq\,0}$ such that 
\begin{equation}
\boldsymbol{\Sigma}_{A\vert X} + t^*\,\boldsymbol{\Pi}_{A\vert X} = (1+t^*)\,\boldsymbol{\Sigma}^{(\rm{inst})}_{A\vert X},
\label{eq:robustness_decomposition_proof}
\end{equation}
for some $\boldsymbol{\Pi}_{A\vert X}\in\mathsf{1SQI}$ and $\boldsymbol{\Sigma}^{(\rm{inst})}_{A\vert X} \in \mathsf{1SQI}$. Applying $\mathcal{M}$ to both sides of this equation gives
\begin{equation}
\mathcal{M}(\boldsymbol{\Sigma}_{A\vert X})+ t^*\,\mathcal{M}(\boldsymbol{\Pi}_{A\vert X}) = (1+t^*)\,\mathcal{M}\left(\boldsymbol{\Sigma}^{(\rm{inst})}_{A\vert X}\right),
\label{eq:robustness_after_map}
\end{equation}
where the linearity of $\mathcal{M}$ has been used. Now, since $\mathcal{M}(\boldsymbol{\Pi}_{A\vert X}),\,  \mathcal{M}(\boldsymbol{\Sigma}^{(\rm{inst})}_{A\vert X})\in\mathsf{1SQI}$ (because  $\mathcal{M}$ is a free operation of QI), Eq. \eqref{eq:robustness_after_map} realises a particular decomposition for $\mathcal{M}(\boldsymbol{\Sigma}_{A\vert X})$ of the form of that of Eq. \eqref{eq:robustness_decomposition_proof} for $\boldsymbol{\Sigma}_{A\vert X}$. Thus, $t^*$ must necessarily be larger or equal than the corresponding minimum for $\mathcal{M}({\Sigma}_{A\vert X})$. That is,
\begin{equation}
R_{\rm{ni}}(\mathcal{M}(\boldsymbol{\Sigma}_{A\vert X})) \leq R_{\rm{ni}}(\boldsymbol{\Sigma}_{A\vert X}),
\label{eq:robustness_monotonicity}
\end{equation}
which proves that $R_{\rm{ni}}$ is a non-instrumentality monotone.

\section{Robustness of non-instrumentality as an SDP optimisation}
\label{sec:Robustness_SDP}
Eq. \eqref{eq:def_robustness} can be re-expressed as $R_{\rm{ni}}(\boldsymbol{\Sigma}_{A\vert X})=t^*$, with $t^*$ defined by Eq. \eqref{eq:robustness_decomposition_proof}. This implies that 
\begin{equation}
\sigma_{a|x}=\,(1+t^*)\sum_{\tilde{\lambda}} D_{\tilde{\lambda}}(a|x)\,\tilde{\sigma}_{a,\tilde{\lambda}} - t^*\,\pi_{a|x},
\label{eq:Robustness_Decomposition}
\end{equation}
with $\boldsymbol{\Pi}_{A\vert X}=\{\pi_{a\vert x}\geq 0\}_{a,\,x}\in\mathsf{1SQI}$. Hence,
\begin{equation}
\pi_{a\vert x} = \sum_{\tilde{\lambda}} D_{\tilde{\lambda}}(a|x)\,\sigma^\prime_{a,\tilde{\lambda}},
\label{eq:1SQInoise_Decomposition}
\end{equation}
for $\sigma^\prime_{a,\tilde{\lambda}} \geq 0$, with $\Tr[\sigma^\prime_{a,\tilde{\lambda}}] = \Tr[\sigma^\prime_{a^\prime,\tilde{\lambda}}],\,\forall\, a,\, a^\prime \in A$. Both $D_{\tilde{\lambda}}(a|x)$ and $\tilde{\sigma}_{a,\tilde{\lambda}}$ are also defined as in Eq. \eqref{eq:Rho_QIDAG_Det}.
The problem of finding $t^*$  is then expressed explicitly as the following SDP:
\begin{subequations}
\label{eq:rob_primal}
\begin{align}
\textrm{Given} &\quad \boldsymbol{\Sigma}_{A\vert X}=\big\{\sigma_{a\vert x}\big\}_{a , x}, \nonumber \\
\label{eq:1SQIrobustness_objective}
\min_{\{\tilde{\chi}_{a,\lambda},\,\chi_{a,\lambda}\in\mathcal{B}(\mathbb{H}_{B})\}_{a, \lambda}}  &\quad t, \\
\label{eq:1SQIrobustness_noisepositivity}
\textrm{s. t.} &\quad\sigma_{a\vert x} = \sum_{\lambda} D_\lambda(a\vert x)\, \left(\tilde{\chi}_{a,\lambda} - \chi_{a,\lambda}\right), \\
\textrm{with} &\quad 1+t= \sum_{\lambda} \Tr\left[\tilde{\chi}_{a,\lambda}\right], \\
\label{eq:1SQIrobustness_localdistribution}
 &\quad \Tr\left[\tilde{\chi}_{a,\lambda}\right]=\,\Tr\left[\tilde{\chi}_{0,\lambda}\right], \\
 \label{eq:1SQIrobustness_localdistribution_noise}
 &\quad \Tr\left[\chi_{a,\lambda}\right]=\,\Tr\left[\chi_{0,\lambda}\right], \\
\label{eq:1SQIrobustness_localstatepositivity}
 &\quad \tilde{\chi}_{a,\lambda} \geq 0, \\
\label{eq:1SQIrobustness_localstatepositivity_noise}
\textrm{and} &\quad \chi_{a,\lambda} \geq 0.
\end{align}
\end{subequations}
Note that normalization of $\sigma_{a|x}$ automatically implies $\sum_\lambda \Tr[\chi_{a,\lambda}] = t$, which explains why one does not impose it as an independent, explicit constraint on the optimization.

Like Eq. \eqref{eq:primal_given}, the test \eqref{eq:rob_primal} admits a dual formulation, which takes the following form.
\begin{subequations}
\label{eq:rob_dual}
\begin{align}
\textrm{Given} &\quad \boldsymbol{\Sigma}_{A\vert X}=\big\{\sigma_{a\vert x}\big\}_{a , x}, \nonumber \\
\label{eq:1SQIrobustnessdual_objective}
\max_{\{V_{a\vert x}\in\mathcal{B}(\mathbb{H}_{B})\}_{a, x}} &\quad \sum_{a,x} \Tr\left[V_{a\vert x}\, \sigma_{a\vert x}\right] - 1, \\
\label{eq:1SQIrobustnessdual_witnesseigenvalue}
\textrm{s. t.} &\quad\ B^\prime_{a,\lambda}\,\openone_B \leq \sum_{x} V_{a\vert x}\, D_\lambda(a\vert x) \leq B_{a,\lambda}\,\openone_B, \\
\label{eq:1SQIrobustnessdual_eigenvaluesum}
\textrm{with} &\quad \sum_{a} B_{a, \lambda}=\,1, \\
\label{eq:1SQIrobustnessdual_lowereigenvaluesum}
\textrm{and} &\quad \sum_{a} B^\prime_{a, \lambda}=\,0.
\end{align}
\end{subequations}
where $B_{a,\lambda},\,B^{\prime}_{a,\lambda}\in\mathbb{R}$ for all $a\in\mathbb{Z}_{A}$ and all $\lambda\in\mathbb{Z}_{\Lambda}$. With the same arguments as in the discussion right after Eqs. \eqref{eq:dual_given}, one sees that Eq. \eqref{eq:1SQIrobustnessdual_objective} returns a positive maximum (the one defining $t^*$) if, and only if, $\boldsymbol{\Sigma}_{A\vert X}\notin\mathsf{1SQI}$. In fact, using that $\boldsymbol{\Sigma}_{A\vert X}$ is well-normalised, the term $-1$ in the objective function can be absorbed into the witness' definition with the variable change $\boldsymbol{V}_{A\vert X}\,\rightarrow\,\boldsymbol{W}_{A\vert X}\coloneqq\{\,V_{a\vert x} - \openone_B/|X|\,\}_{a,x}$. The resulting SDP (for the redefined witness $\boldsymbol{W}_{A\vert X}$) is similar to the one in Eqs. \eqref{eq:dual_given}, but with an extra constraint coming from the left-hand side inequality of Eq. \eqref{eq:1SQIrobustnessdual_witnesseigenvalue}, and with the witness scale no longer fixed. Thus, the robustness is given by the violation of the optimal $\boldsymbol{V}_{A\vert X}$ over all non-instrumentality witnesses with 1SQI bound $\beta=1$ and subject to the specific constraints given by Eqs. \eqref{eq:rob_dual}.

\section{Proof of theorem \ref{theo:two}}
\label{sec:proof_main_theo}
Consider an arbitrary $\boldsymbol{\Sigma}^{(\rm{qinst})}_{A\vert X} \in \mathsf{QI}$. By definition, it admits a decomposition as in Eq.  \eqref{eq:assemblage_qi}. Here, we use the short-hand notation $\boldsymbol{\Sigma}^{(\rm{qinst})}_{A\vert X} = \boldsymbol{\mathcal{E}}_{B\vert A,\Lambda_B}\big(\boldsymbol{\Sigma}^{(\rm{q})}_{A\vert X}\big)$, where $\boldsymbol{\mathcal{E}}_{B\vert A,\Lambda_B}\coloneqq\{\mathcal{E}_{B\vert a,\Lambda_B}\}_a$, to represent Eq.  \eqref{eq:assemblage_qi}. In addition, we denote by $\boldsymbol{\mathcal{E}}^*_{B\vert A,\Lambda_B}\coloneqq\{\mathcal{E}^*_{B\vert a,\Lambda_B}\}_a$ the collection of dual (adjoint) maps $\mathcal{E}^*_{B\vert a,\Lambda_B}$ of each completely-positive trace-preserving (CPTP) map $\mathcal{E}_{B\vert a,\Lambda_B}$. This has the property that $\big\langle \boldsymbol{V}_{A\vert X}, \boldsymbol{\mathcal{E}}_{B\vert A,\Lambda_B}\big(\boldsymbol{\Sigma}_{A\vert X}\big)\big\rangle =\big\langle \boldsymbol{\mathcal{E}}^*_{B\vert A,\Lambda_B}\big(\boldsymbol{V}_{A\vert X}\big), \boldsymbol{\Sigma}_{A\vert X}\big\rangle$, for any $\boldsymbol{V}_{A\vert X}$ and $\boldsymbol{\Sigma}_{A\vert X}$.
Then, if $\boldsymbol{V}_{A\vert X}$ is the optimal robustness witness of $\boldsymbol{\Sigma}^{(\rm{qinst})}_{A\vert X}$ , defined by Eqs. \eqref{eq:rob_dual}, it holds that
\begin{align}
R_{\rm{ni}}\left(\boldsymbol{\Sigma}^{(\rm{qinst})}_{A\vert X}\right)&=\left\langle \boldsymbol{V}_{A\vert X}, \boldsymbol{\Sigma}^{(\rm{qinst})}_{A\vert X}\right\rangle - 1 \nonumber \\
&=\left\langle \boldsymbol{V}_{A\vert X}, \boldsymbol{\mathcal{E}}_{B\vert A,\Lambda_B}\left(\boldsymbol{\Sigma}^{(\rm{q})}_{A\vert X}\right)\right\rangle - 1 \nonumber \\
&=\left\langle \boldsymbol{\mathcal{E}}^*_{B\vert A,\Lambda_B}\left(\boldsymbol{V}_{A\vert X}\right), \boldsymbol{\Sigma}^{(\rm{q})}_{A\vert X}\right\rangle - 1.
\label{eq:ThmIIproof_partI}
\end{align}

Now, assume, for a moment, that $\boldsymbol{\mathcal{E}}^*_{B\vert A,\Lambda_B}\left(\boldsymbol{V}_{A\vert X}\right)$ is also a valid robustness witness. Then, denoting by  $\boldsymbol{U}_{A\vert X}$ the optimal robustness witness for $\boldsymbol{\Sigma}^{(\rm{q})}_{A\vert X}$, it must hold that
\begin{equation}
\left\langle\boldsymbol{\mathcal{E}}^*_{B\vert A,\Lambda_B}\left(\boldsymbol{V}_{A\vert X}\right), \boldsymbol{\Sigma}^{(\rm{q})}_{A\vert X}\right\rangle - 1 \leq \left\langle \boldsymbol{U}_{A\vert X}, \boldsymbol{\Sigma}^{(\rm{q})}_{A\vert X}\right\rangle - 1.
\label{eq:ThmIIproof_partII}
\end{equation}
The left-hand side of this equation equals $R_{\rm{ni}}\left(\boldsymbol{\Sigma}^{(\rm{qinst})}_{A\vert X}\right)$, whereas the right-hand side equals $R_{\rm{ni}}\left(\boldsymbol{\Sigma}^{(\rm{q})}_{A\vert X}\right)$, thus giving $R_{\rm{ni}}\left(\boldsymbol{\Sigma}^{(\rm{q})}_{A\vert X}\right) \geq R_{\rm{ni}}\left(\boldsymbol{\Sigma}^{(\rm{qinst})}_{A\vert X}\right)$. So, the only missing ingredient is to show that $\boldsymbol{\mathcal{E}}^*_{B\vert A,\Lambda_B}\left(\boldsymbol{V}_{A\vert X}\right)$ is, in fact, a valid robustness witness.

To prove this we note that, for any $\boldsymbol{\Sigma}^{(\rm{inst})}_{A\vert X}\in \mathsf{1SQI}$, $\boldsymbol{\mathcal{E}}_{B\vert A,\Lambda_B}(\boldsymbol{\Sigma}^{(\rm{inst})}_{A\vert X})$ is also in $\mathsf{1SQI}$. Hence, $\left\langle\boldsymbol{\mathcal{E}}^*_{B\vert A,\Lambda_B}\left(\boldsymbol{V}_{A\vert X}\right), \boldsymbol{\Sigma}^{(\rm{inst})}_{A\vert X}\right\rangle = \left\langle \boldsymbol{V}_{A\vert X}, \boldsymbol{\mathcal{E}}_{B\vert A,\Lambda_B}\left(\boldsymbol{\Sigma}^{(\rm{inst})}_{A\vert X}\right)\right\rangle \leq 1$, for all $\boldsymbol{\Sigma}^{(\rm{inst})}_{A\vert X}\in \mathsf{1SQI}$. This implies that $\boldsymbol{\mathcal{E}}^*_{B\vert A,\Lambda_B}\left(\boldsymbol{V}_{A\vert X}\right)$ is a non-instrumentality witness with $\beta_{\boldsymbol{\mathcal{E}}^*_{B\vert A,\Lambda_B}\left(\boldsymbol{V}_{A\vert X}\right)}=1$. Also, given that each $\mathcal{E}^*_{B\vert a,\Lambda_B}$ is completely-positive (CP) and unital, since it is the dual of a CPTP map, applying these dual maps to any robustness witness does not invalidate its defining constraints, in Eqs. \eqref{eq:1SQIrobustnessdual_witnesseigenvalue}-\eqref{eq:1SQIrobustnessdual_lowereigenvaluesum}. This means that the resulting object after the application of the dual map is also a valid 1SQI-robustness witness in the SDP formulation.
$\qed$


\begin{thebibliography}{68}
\bibitem [{Thi()}]{ThirdComment}%
  \BibitemOpen
  \href@noop {} {}\bibinfo {note} {Importantly, node $A$ is kept classical
  throughout. The term ``quantum" in QI is used here as a short form to refer
  to ``quantum common-cause". Thus, QI should not be mistaken with the fully
  quantum case where all nodes, including $A$, are quantised.}\BibitemShut
  {Stop}%
\bibitem [{Fir()}]{FirstComment}%
  \BibitemOpen
  \href@noop {} {}\bibinfo {note} {Note that, interestingly, instrumental
  causal models display the same DAG for both the DI and the 1-sided DI cases.
  This is not the case for LHV and LHS models. The former need an extra (input)
  node with respect to the latter.}\BibitemShut {Stop}%
\bibitem [{Sec()}]{SecondComment}%
  \BibitemOpen
  \href@noop {} {}\bibinfo {note} {Interestingly, optimising the violation of a
  fixed NI witness over all quantum assemblages can -- thanks to the
  no-signalling constraint -- be directly recast as an SDP problem. In
  contrast, this is not true for QI assemblages.}\BibitemShut {Stop}%
\end{thebibliography}
\end{document}